\documentclass[aps,prb,twocolumn]{revtex4}

\usepackage{graphics}
\usepackage{amsmath}
\newcommand\includefigure[2]{\centerline{%
  \resizebox{#1}{!}{\includegraphics{#2}}%
}}
\newcommand\bm[1]{\text{\boldmath $#1$}}

\begin{document}

\title{Universal Reduction of Effective Coordination Number \\ in
the Quasi-One-Dimensional Ising Model}

\author{Synge Todo}
\email{wistaria@ap.t.u-tokyo.ac.jp}

\affiliation{Department of Applied Physics, University of Tokyo, Tokyo
  113-8656, Japan}
\affiliation{CREST, Japan Science and Technology Agency, Kawaguchi,
  332-0012, Japan}

\begin{abstract}
 Critical temperature of quasi-one-dimensional general-spin Ising
 ferromagnets is investigated by means of the cluster Monte Carlo method
 performed on infinite-length strips, $L \times \infty$ or $L \times L
 \times \infty$.  We find that in the weak interchain coupling regime
 the critical temperature as a function of the interchain coupling is
 well-described by a chain mean-field formula with a reduced effective
 coordination number, as the quantum Heisenberg antiferromagnets
 recently reported by Yasuda \textit{et al.} [Phys. Rev. Lett. {\bf 94}, 
 217201 (2005)].  It is also confirmed that the effective coordination number
 is independent of the spin size.  We show that in the
 weak interchain coupling limit the effective coordination number is,
 irrespective of the spin size, rigorously given by the quantum critical
 point of a spin-1/2 transverse-field Ising model.
\end{abstract}

\date{$ $Id: main.tex,v 1.31 2006/09/12 08:23:56 wistaria Exp $ $\!\!}

\maketitle

\section{Introduction}
\label{sec:intro}

Effects of interchain coupling in quasi-one-dimensional (Q1D) magnets
have been extensively investigated over many years.  Theoretically,
introduction of an infinitesimal interchain coupling can bring a drastic
change to the system; emergence of a thermal phase transition and
long-range magnetic order therewith at finite temperatures.  In
interpreting experimental results on Q1D materials, which mostly exhibit
anomalous behavior at finite temperatures, one often has to take the
interchain effects into account as well.  A standard and established
approach for incorporating such a weak interchain interaction is the
chain mean-field (CMF) approximation,\cite{ScalapinoIP1975, Schulz1996}
in which a problem in two or three dimensions is reduced to a
single-chain model in an effective field by ignoring fluctuations of
order parameter between weakly coupled chains.  Within the CMF theory,
the magnetic susceptibility of a Q1D system is approximated as
\begin{equation}
 \label{eqn:cmf-chi}
  \chi(T) = \frac{\chi_\text{1d}(T)}{1 - zJ' \chi_\text{1d}(T)},
\end{equation}
where $\chi_\text{1d}(T)$ is the susceptibility of a {\em genuinely
one-dimensional} chain, $z$ the coordination number, i.e., the
number of nearest neighbor sites in directions perpendicular to the
chain, and $J'$ is the interchain coupling constant in these directions.
Since $\chi_\text{1d}(T)$ may not have any singularity at finite
temperatures, the critical temperature $T_\text{c}$ of the Q1D system is
given as the locus of the simple pole in the right-hand side (rhs) of
Eq.~(\ref{eqn:cmf-chi}).  That is,
\begin{equation}
 \label{eqn:cmf-tc} z = \frac{1}{J' \chi_\text{1d}(T_\text{c})}
\end{equation}
is the equation which determines the Q1D critical temperature.  It has been
confirmed that for various Q1D (and also Q2D) materials the solution of
the CMF self-consistent equation~(\ref{eqn:cmf-tc}) reproduces a
qualitatively acceptable $J'$-dependence of the critical temperature in
the weak interchain coupling regime.

Recently, Yasuda \textit{et al.}\cite{YasudaTHAKTT2005} carefully reexamined the
N\'eel temperature of the quantum Heisenberg antiferromagnets on a
anisotropic cubic lattice, and found that Eq.~(\ref{eqn:cmf-tc}) does
not hold even in the weak interchain coupling limit. In
Ref.~\onlinecite{YasudaTHAKTT2005}, the rhs of Eq.~(\ref{eqn:cmf-tc})
for various $J'$'s is evaluated accurately by the extensive quantum Monte
Carlo simulation using the continuous-imaginary-time loop algorithm, and
it is concluded that for $J' \rightarrow 0$ it converges not to 4 but to
2.78 for the Heisenberg antiferromagnets on the Q1D cubic lattice.  This
discrepancy with the CMF theory is interpreted as a reduction of the
{\em effective coordination number}
\begin{equation}
 \label{eqn:def-zeta}
 \zeta(J') \equiv
 \frac{1}{J'\chi_\text{1d}(T_\text{c}(J'))}
\end{equation}
from that of the original lattice, $z$.  Furthermore, the numerical data
presented in Ref.~\onlinecite{YasudaTHAKTT2005} clearly demonstrate
that the renormalized coordination number is ``universal,'' i.e.,
independent of the spin size $S$ (including the classical limit
$S=\infty$).  Although there have been several
attempts\cite{IrkhinK1997, IrkhinK1998, IrkhinKK1999, IrkhinK2000,
Bocquet2002, HastingsM2006, PrazMH0606032, YaoS0606341} to develop
theories beyond the CMF for specific models so far, our understanding of
$S$-independent renormalization of the effective coordination number is,
fairly speaking, not yet satisfactory.

Another interesting question is whether and how such a renormalization
of the effective coordination number is observed for other
models than the Heisenberg antiferromagnets.  Among many theoretical
spin models, a linear array of $S=1/2$ Ising chains, which forms an
anisotropic square lattice, provides a valuable example, since the exact
critical temperature is known for any ratio of $J'$ to the intrachain
coupling $J$.  Indeed it was pointed out more than three decades
ago\cite{ScalapinoIP1975} that the CMF result for the critical
temperature
\begin{equation}
 \label{eqn:asym-tc-cmf}
 T_\text{c} \simeq \frac{J}{2 \ln (J/J') - 2 \ln \ln (J/J')}
\end{equation}
for $J'/J \ll 1$ is not compatible with the {\em exact} asymptotic
behavior
\begin{equation}
 \label{eqn:asym-tc-exact}
 T_\text{c} \simeq \frac{J}{2 \ln (2J/J') - 2 \ln \ln (2J/J')}.
\end{equation}
Here and hereafter we put the Boltzmann constant to unity.  The former
asymptotic expression is obtained from Eq.~(\ref{eqn:cmf-tc}) with
$z=2$, combined with the exact susceptibility of an $S=1/2$ single Ising
chain
\begin{equation}
 \label{eqn:chi-half}
 \chi_\text{1d}^{1/2}(T) = \frac{1}{4 T} e^{J/2 T},
\end{equation}
whereas the latter is obtained by expanding Onsager's exact solution for
the anisotropic Ising model\cite{Onsager1944}
\begin{equation}
 \label{eqn:onsager}
 \sinh \frac{J}{2 T_\text{c}} \sinh \frac{J'}{2 T_\text{c}} = 1
\end{equation}
for small $J'/J$.  [Note that here the absolute spin length is not 1 but
1/2.  See the definition of our Hamiltonian~(\ref{eqn:hamiltonian})
below.]  Interestingly, if the effective coordination number
[Eq.~(\ref{eqn:def-zeta})] is evaluated for this case by using the exact
critical temperature and the exact one-dimensional susceptibility,
Eqs.~(\ref{eqn:chi-half}) and (\ref{eqn:onsager}) respectively, one
immediately finds that it converges to unity, instead of 2, in the $J'/J
= 0$ limit.\cite{PrazMH0606032} This example thus clearly demonstrates
the breakdown of the CMF approach for Q1D magnets, to which we will
address ourselves in the present work.

In the present paper, we report our detailed analysis on the Q1D
general-spin Ising models, and discuss the origin of the $S$-independent
renormalization in
the effective coordination number in the weak interchain coupling
regime.  The organization of the paper is as follows: In
Sec.~\ref{sec:model}, we define our models and summarize some basic
properties in the genuinely one-dimensional case.  In
Sec.~\ref{sec:method}, we describe our Monte Carlo method and the detail
of finite-size scaling analysis.  The main results of our Monte Carlo
simulation are presented in Sec.~\ref{sec:results}, where we
unambiguously show that the critical temperature is described quite well
by a CMF formula with a reduced effective coordination number.  The
effective coordination number is found independent of $S$
in the weak interchain coupling limit, but does depend on the dimensionality of the
lattice as well as on the fine structure of the lattice in directions
perpendicular to the chains.  In Sec.~\ref{sec:mapping}, we discuss the
origin of the ``universality'' (or ``nonuniversality'') by considering a
mapping to a quantum Ising model in the weak interchain coupling limit.
We show that the original spin-$S$ Ising model is mapped onto the
spin-$1/2$ transverse-field Ising model, irrespective of the spin size,
and furthermore the effective coordination number in this limit is
rigorously independent of $S$ and given by the quantum critical point of
the mapped $S=1/2$ transverse-field Ising model.  Our results and
conclusions are summarized in Sec.~\ref{sec:summary}.

\section{Quasi-one-dimensional Ising Model}
\label{sec:model}

\begin{figure}
 \includefigure{0.44\textwidth}{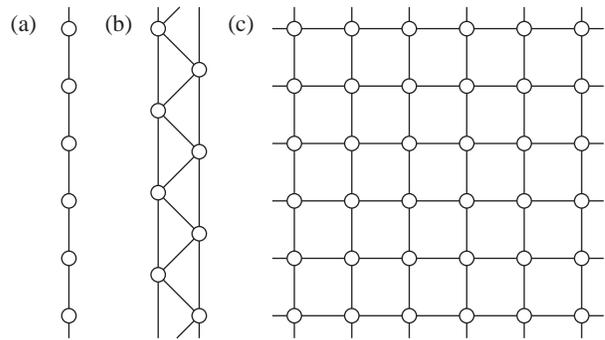}
 \caption{Lattice structures in the transverse directions perpendicular
 to the chains: (a) square lattice (SQ), (b) square lattice with
 anisotropic next-nearest-neighbor interaction (SQ+ANNNI), and (c)
 simple cubic lattice (SC).  All the bonds shown in the figure have the
 exchange coupling of the same strength, $J'$.}
 \label{fig:lattices}
\end{figure}

We consider the following classical spin-$S$ Ising model:
\begin{equation}
 \label{eqn:hamiltonian}
 {\cal H} = -J \sum_{i,\bm{r}} S_{i,\bm{r}}^z S_{i+1,\bm{r}}^z - J'
  \sum_{i,\bm{r}, \bm{\delta}} S_{i,\bm{r}}^z S_{i,\bm{r}+\bm{\delta}}^z ,
\end{equation}
where $i$ and $\bm{r}$ are the position of lattice sites along chains
and perpendicular directions, respectively, $\bm{\delta}$ is summed over
the $z/2$ nearest-neighbor vectors in the transverse directions, and
$S_{i,\bm{r}}^z$ $(= S,S-1,\ldots,-S)$ is the spin variable at site
$(i,\bm{r})$.  Both of the longitudinal and transverse coupling
constants, $J$ and $J'$, are assumed to be ferromagnetic ($J \ge J' >
0$).

In the present study, three different Q1D lattices are considered:
square lattice (SQ), square lattice with anisotropic
next-nearest-neighbor interaction (SQ+ANNNI), and simple cubic lattice
(SC).  In Fig.~\ref{fig:lattices}, the lattice structure in a slice
perpendicular to the chains are drawn.  The former two (SQ and SQ+ANNNI)
are two-dimensional lattices, while the last one (SC) is three
dimensions.  On the other hand, the latter two shares the same
coordination number, $z=4$, which is larger than that of the first
lattice ($z=2$).  By considering these three lattices, one can evaluate
separately the effects of dimensionality from the coordination number.

The linear dimension of the lattice in the directions perpendicular to
the chains is specified by $L$.  We assume periodic boundary conditions
in these directions.  For the SQ and SQ+ANNNI lattices, the number of
spins in each slice, denoted by $N_\text{s}$ hereafter, is equal to $L$,
while it is $L^2$ for the SC lattice.  In the chain direction, on the
other hand, the system size is assumed to be infinite as explained in
the next section.

Before starting a detailed discussion about the critical temperature of
the Q1D system, let us briefly review thermal properties of a genuinely
one-dimensional single chain, i.e., Eq.~(\ref{eqn:hamiltonian}) with
$J'=0$, since its susceptibility $\chi_\text{1d}(T)$ appears as a key
quantity in the CMF approximation, and thus is expected to dominate
asymptotic behavior of the critical temperature $T_\text{c}(J')$ for
$J' \ll J$.

The partition function of a periodic spin-$S$ Ising chain of length $M$
is represented as\cite{KramersW1941}
\begin{equation}
 \label{eqn:1d-z}
  Z_\text{1d} = \text{tr} \, \bm{T}^M.
\end{equation}
Here the transfer matrix $\bm{T}$ is a $(2S+1)\times(2S+1)$ real
symmetric matrix with matrix elements
\begin{equation}
 \bm{T}_{ij} = e^{J (S+1-i) (S+1-j) / T}.
\end{equation}
The matrix $\bm T$ is diagonalized as
\begin{equation}
 \label{eqn:diagonalization}
 \bm{T} = \bm{U} \bm{D} \bm{U}^{-1}
\end{equation}
by a diagonal matrix of eigenvalues, $\bm{D}_{ij} = \lambda_{i}
\delta_{i,j}$, and a matrix of the corresponding right (column)
eigenvectors, $\bm{U} = (u_1,u_2,\ldots,u_{2S+1})$ with $\bm{T} u_i =
\lambda_i u_i$.  We assume all the eigenvectors are normalized, i.e.,
$\bm{U}^{-1} = \, ^t\bm{U}$.  Without loss of generality we further
assume $\lambda_1$ and $\lambda_2$ denote the largest and the second
largest eigenvalues, respectively.

At finite temperatures, the transfer matrix $\bm{T}$ is positive
definite and its largest eigenvalue $\lambda_1$ is nondegenerate.  Thus
the partition function~(\ref{eqn:1d-z}) is evaluated as $(\lambda_1)^M$
in the thermodynamic limit.  Similarly, the thermal average of
correlation function between the spins at site 0 and $r$ is calculated
in terms of the transfer matrix as 
\begin{equation}
 \begin{split}
  \langle S^z_0 S^z_r \rangle &= \frac{1}{Z} \text{tr} \, \bm{S}
  \bm{T}^r \bm{S}
  \bm{T}^{M-r} \\
  &= \frac{1}{Z} \text{tr} \, (\bm{U}^{-1} \bm{S} \bm{U}) \bm{D}^r
  (\bm{U}^{-1} \bm{S} \bm{U})
  \bm{D}^{M-r} \\
  & \!\!\!\! \underset{M\rightarrow\infty}{\longrightarrow}
  \lambda_1^{-r} \text{tr} \, (\bm{U}^{-1} \bm{S} \bm{U}) \bm{D}^r
  (\bm{U}^{-1} \bm{S} \bm{U}) \\
  &= \sum_{i=2}^{2S+1} |^t u_1 \bm{S} u_i|^2
  \Big( \frac{\lambda_i}{\lambda_1} \Big)^r,
 \end{split}
\end{equation}
where we introduce $\bm{S}_{ij} = (S+1-i)\delta_{i,j}$.  In the last
expression, we explicitly dropped $i=1$ in the summation, since
$(^tu_1\bm{S}u_1)$ constantly vanishes due to the spin inversion
symmetry in the Hamiltonian~(\ref{eqn:hamiltonian}).  The susceptibility
is then obtained by summing up the correlation function over the whole
lattice
\begin{equation}
 \label{eqn:exact-chi-1d}
 \begin{split}
  \chi_\text{1d}(T) &= \frac{1}{T} \sum_{r=-\infty}^{\infty}\langle
  S^z_0 S^z_r \rangle 
  = \frac{1}{T} \sum_{i=2}^{2S+1} |^tu_1 \bm{S} u_i|^2
  \frac{\lambda_1+\lambda_i}{\lambda_1-\lambda_i}.
 \end{split}
\end{equation}
For $S=1/2$, 1, and $3/2$, all the eigenvalues and the eigenvectors of
the transfer matrix can be obtained explicitly in a closed form, and so
is the susceptibility.\cite{SuzukiTK1967} For larger $S$, it is not able
to solve the eigenvalue problem analytically, but one can still
diagonalize $\bm{T}$ numerically very easily.

\begin{figure}
 \includefigure{0.44\textwidth}{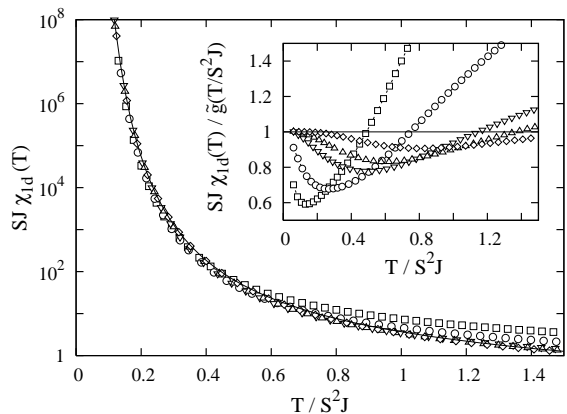}
 \caption{Scaling
 plot of susceptibility of a single chain with $S=1/2$ (solid
 line), 1 (diamonds), 3/2 (upward triangles), 2 (downward triangles), 4
 (circles), and 8 (squares).  In the inset, corrections to the
 low-temperature asymptotic form [Eq.~(\ref{eqn:asym-chi-1d})] are
 presented.}  \label{fig:chi-1d}
\end{figure}

At very low temperatures, $T \ll J$, the largest two eigenvalues
$\lambda_1$ and $\lambda_2$ nearly degenerate with each other, which
reflects two degenerating fully polarized ground states,
$(S,S,\ldots,S)$ and its spin-inverted counterpart, with energy
$-S^2JM$.  The corresponding eigenvectors are symmetric or
antisymmetric, respectively, under the spin inversion, and their
elements are given as follows:
\begin{align}
 \label{eqn:14}
 u_{1,1} &= u_{1,2S+1} = \frac{1}{\sqrt{2}} + O(e^{-J/T}), \\
 \label{eqn:15}
 u_{1,i} &= O(e^{-J/T}) \ \ \text{for $i=2,\ldots,2S$,} \\
 \label{eqn:16}
 u_{2,1} &= -u_{1,2S+1} = \frac{1}{\sqrt{2}} + O(e^{-J/T}), \\
 \label{eqn:17}
 u_{2,i} &= O(e^{-J/T}) \ \ \text{for $i=2,\ldots,2S$}.
\end{align}
The absolute value of the subdominant eigenvalues is exponentially small
in comparison with the largest two.  The correlation function and the
susceptibility at low temperatures are thus written as
\begin{align}
 \langle S^z_0 S^z_r \rangle &\simeq S^2
 \Big( \frac{\lambda_2}{\lambda_1} \Big)^r \ \ \text{and} \\
 \chi_\text{1d}(T) &\simeq \frac{S^2}{T}
  \frac{\lambda_1+\lambda_2}{\lambda_1-\lambda_2},
\end{align}
respectively.  By defining the correlation length $\xi_\text{1d}(T)$ as
$\xi_\text{1d}^{-1}(T) = -\log (\lambda_2/\lambda_1)$, we obtain
\begin{align}
 \label{eqn:20}
 \chi_\text{1d}(T) &\simeq
 \frac{S^2}{T}
 \frac{1+e^{-1/\xi_\text{1d}(T)}}{1-e^{-1/\xi_\text{1d}(T)}}
 \simeq
 \frac{2S^2}{T} \xi_\text{1d}(T)
\end{align}
for general spin $S$.  For $S=1$ and 3/2, the low-temperature
susceptibility is explicitly given by\cite{SuzukiTK1967}
\begin{align}
 \label{eqn:22}
 \chi_\text{1d}^{1}(T) &\simeq \frac{1}{2T} e^{2J/T} \ \ \text{and} \\
 \label{eqn:23}
 \chi_\text{1d}^{3/2}(T) &\simeq \frac{3}{4T} e^{9J/2T},
\end{align}
respectively. [The exact expression for $S=1/2$ is already presented in
Eq.~(\ref{eqn:chi-half}).]  For larger $S$, analytic expressions are not
known, but low-temperature asymptotic forms for the correlation length
and the susceptibility are conjectured as follows:
\begin{align}
 \label{eqn:asym-xi-1d}
 \begin{split}
  \xi_\text{1d}^S(T) &\simeq 
  - \Big[ 2S \log \Big( \tanh \frac{S^2J}{T} \Big) \Big]^{-1} \\
  & \simeq
  \frac{1}{4S} e^{2S^2J/T} = 
  \frac{1}{S} \tilde{f}(T/S^2J) 
 \end{split}
\end{align}
and
\begin{align}
 \label{eqn:asym-chi-1d}
 \chi_\text{1d}^S(T) &\simeq 
 \frac{S}{2T} e^{2S^2J/T} = \frac{1}{SJ} \tilde{g}(T/S^2J),
\end{align}
where $\tilde{f}(x) = e^{2/x} / 4$ and $\tilde{g}(x) = e^{2/x} / 2x$ are
scaling functions.  These expressions are confirmed
explicitly, e.g., Eqs.~(\ref{eqn:chi-half}), (\ref{eqn:22}), and
(\ref{eqn:23}), for $S \le 3/2$.  We emphasize that these scaling forms
are {\em not} very trivial.  Indeed, a naive dimensional analysis yields
$\xi_\text{1d}^S(T) \simeq \tilde{f}(T/S^2J)$ and $\chi_\text{1d}^S(T)
\simeq J^{-1} \tilde{g}(T/S^2J)$ with some scaling functions
$\tilde{f}(x)$ and $\tilde{g}(x)$, which differs from the correct
scaling forms presented above by a factor $1/S$.  In
Fig.~\ref{fig:chi-1d}, we present a scaling plot of the numerically
exact data for the susceptibility, given by
Eq.~(\ref{eqn:exact-chi-1d}), for $S=1/2,1,\ldots,8$, which confirms the
asymptotic scaling formula~(\ref{eqn:asym-chi-1d}).

\section{Numerical Method}
\label{sec:method}

\subsection{Cluster Monte Carlo on infinite-length strip}

For the $S=1/2$ SQ lattice, the critical temperature is solved exactly
for any combination of $J'$ and $/J$ as we have already seen in
Sec.~\ref{sec:intro}.  However, no exact solutions are available for $S
\ge 1$.  For the other lattices (SQ+ANNNI and SC) there are no solutions
even for $S=1/2$.

In the present study we employ the Monte Carlo method combined with the
finite-size scaling analysis for evaluating unbiased critical
temperatures of the Q1D Ising models.  Since the present model has only
the ferromagnetic couplings and has no frustration, the Swendsen-Wang
cluster algorithm,\cite{SwendsenW1987} or the Wolff cluster
algorithm,\cite{Wolff1989} is one of the best algorithms.  To optimize
further for the Q1D general-spin Ising models, we extend the cluster
Monte Carlo algorithm in the following two points.

The first optimization is for strong anisotropy.  Since our model is
strongly spatially anisotropic, one has to fine-tune the aspect ratio of
the lattice to simulate an effectively large system with a reasonable
cost.\cite{YasudaTHAKTT2005, MatsumotoYTT2001} In the present study,
instead of anisotropic finite lattices, we simulate infinite length
strips $N_\text{s} \times \infty$ directly, following the
infinite-lattice algorithm proposed by Evertz and von der
Linden.\cite{EvertzL2001} In the original proposal of the
infinite-lattice cluster algorithm, the lattice is assumed to be
infinite in all the spatial directions and only one cluster including
the origin is build at each Monte Carlo step.  In our simulation,
however, the lattice is infinite only in one direction, the chain
direction, which has a stronger coupling than the transverse directions,
and we build up completely all clusters including the $N_\text{s}$ sites on
the slice in the center of the lattice at each Monte Carlo step.  This
modification avoids the divergence of cluster size and enables us the
simulation at and below the critical temperature of the bulk and thus
the finite-size scaling for accurate estimation the critical
temperature.

Another modification to the original cluster algorithm is the extension
for general spin $S > 1/2$.\cite{TodoK2000, TodoK2001} To
this end, first we decompose a spin into $2S$ {\em subspins}, that is,
we simulate the following $S=1/2$ model
\begin{equation}
 \label{eqn:subspin}
 \begin{split}
  {\cal H}_\text{subspin} =& -J \sum_{i,\bm{r}} \sum_{\alpha,
  \alpha'=1}^{2S} \tilde{S}_{i,\bm{r},\alpha}^z
  \tilde{S}_{i+1,\bm{r},\alpha'}^z \\ &
  - J'
  \sum_{i,\bm{r},\bm{\delta}} \sum_{\alpha,\alpha'=1}^{2S}
  \tilde{S}_{i,\bm{r},\alpha}^z
  \tilde{S}_{i,\bm{r}+\bm{\delta},\alpha'}^z,
 \end{split}
\end{equation}
instead of the original spin-$S$ model.  Here
$\tilde{S}_{i,\bm{r},\alpha} = \pm 1/2$ is the $\alpha$th subspin at
site $(i,\bm{r})$.  The volume of the phase space of the mapped
system~(\ref{eqn:subspin}) is enlarged than the original one [from
$(2S+1)^N$ to $2^{2SN}$, where $N$ is the number of original spins].  We
need an extra procedure which symmetrize each spin to recover the
original phase space\cite{TodoK2000, TodoK2001} as described below.

A Monte Carlo step is as follows: (i) set $i \mapsto 0$; (ii) every
possible subspin pair, $\tilde{S}_{i,\bm{r},\alpha}$ and
$\tilde{S}_{i,\bm{r+\delta},\alpha'}$, in the $i$th slice is connected
with probability $(1-e^{-2J'/T})$ if $\tilde{S}_{i,\bm{r},\alpha} =
\tilde{S}_{i,\bm{r+\delta},\alpha'}$, or left disconnected otherwise;
(iii) for each group of subspins
$(i,\bm{r})$, the subspins are partitioned into two according to their
spin direction, a random permutation is generated for each part, and the
subspins belonging to the same permutation cycle are connected with
probability one; (iv) every possible subspin pair,
$\tilde{S}_{i,\bm{r},\alpha}$ and $\tilde{S}_{i+1,\bm{r},\alpha'}$, on
the neighboring slices, $i$ and $i+1$, is connected with probability
$(1-e^{-2J/T})$ if $\tilde{S}_{i,\bm{r},\alpha} =
\tilde{S}_{i+1,\bm{r},\alpha'}$, or left disconnected otherwise;
(v) increase $i$ by 1 and repeat
(ii)--(iv) until no clusters in the current slice include any subspins
on the central slice at $i=0$; (vi) repeat (ii)--(v) in the opposite chain
direction, $i=-1,-2,\ldots$; (vii) generate a new spin configuration by
flipping all subspins on each cluster including the subspins on the
central slice randomly with probability $1/2$, and discard all the
clusters.

The present method satisfies detailed balance conditions and ergodicity
for any finite region including the central slice.  Typically, we spend
$1.0 \times 10^6$ Monte Carlo steps for measurement after discarding
$3.3 \times 10^4$ steps for thermalization for each parameter set
$(J'/J, T/J, L)$.

\subsection{Susceptibility and correlation length}

Correlation functions $\langle S^z_{i,\bm{r}} S^z_{i',\bm{r}'} \rangle$
can be measured efficiently in terms of a neat technique, so called
{\em improved estimator},\cite{Wolff1989} i.e.,
\begin{equation}
 \label{eqn:correlation}
 \langle S^z_{i,\bm{r}} S^z_{i',\bm{r}'}
  \rangle = \frac{1}{4} \sum_{\alpha,\alpha'} \langle C(i,\bm{r},\alpha;
  i',\bm{r}',\alpha') \rangle_\text{MC},
\end{equation}
where $\langle \cdots \rangle_\text{MC}$ denotes a Monte Carlo average
and $C(i,\bm{r},\alpha; i',\bm{r}',\alpha')$ takes 1 if the subspins at
$(i,\bm{r},\alpha)$ and $(i',\bm{r}',\alpha')$ belong to the same cluster
and takes 0 otherwise at each Monte Carlo step.  We can further rewrite
Eq.~(\ref{eqn:correlation}) by introducing a function
$C_\ell(i,\bm{r},\alpha)$, which takes 1 if the subspin
$\tilde{S}^z_{i,\bm{r},\alpha}$ is on the $\ell$th cluster and takes 0
otherwise, as
\begin{equation}
 \label{eqn:correlation2}
 \langle S^z_{i,\bm{r}} S^z_{i',\bm{r}'}
  \rangle = \frac{1}{4} \sum_{\alpha,\alpha'} \Big\langle
  \sum_{\ell} C_\ell(i,\bm{r},\alpha) C_\ell(i',\bm{r}',\alpha')
  \Big\rangle_\text{MC}.
\end{equation}

Here one must recall that in the present algorithm clusters will be
built only for a part of the lattice, and there is no way to judge
whether two subspins belong to the same cluster when neither is on the
built clusters.  Therefore the above improved estimator for the
correlation function is valid only if at least one of
$(i,\bm{r},\alpha)$ and $(i',\bm{r}',\alpha')$ is guaranteed to belong
always to built clusters.  Since only the subspins on the central slice
satisfies this condition, $i$ or $i'$ must be zero in
Eqs.~(\ref{eqn:correlation}) and (\ref{eqn:correlation2}).

The magnetic susceptibility $\chi$ is obtained by integrating the
correlation function over the whole lattice.  Fixing $i=0$ in
Eq.~(\ref{eqn:correlation}) and taking a sum over all the other indices,
we obtain the following expression for the susceptibility:
\begin{equation}
 \begin{split}
  \chi &= \frac{1}{TN_\text{s}} \sum_{i=-\infty}^{\infty} \sum_{\bm{r},\bm{r}'}
  \langle S^z_{0,\bm{r}} S^z_{i,\bm{r}'}
  \rangle \\
  &= \frac{1}{4TN_\text{s}} \sum_{i=-\infty}^{\infty}
  \sum_{\bm{r},\bm{r}',\alpha,\alpha'} 
  \Big\langle \sum_\ell C_\ell(0,\bm{r},\alpha) C_\ell(i,\bm{r}',\alpha')
  \Big\rangle_\text{MC} \\
  &= \frac{1}{4TN_\text{s}} \Big\langle \sum_{\ell} n_\ell^{(0)} n_\ell
  \Big\rangle_\text{MC},
 \end{split}
\end{equation}
where $n_\ell^{(0)} \equiv \sum_{\bm{r},\alpha} C_\ell(0,\bm{r},\alpha)$
denotes the number of subspins on the central slice in the $\ell$th
cluster, and $n_\ell \equiv \sum_{i,\bm{r},\alpha}
C_\ell(i,\bm{r},\alpha)$ is the number of subspins in the $\ell$th
cluster.

The correlation length along the strip is estimated as
\begin{equation}
 \xi_{k,k'} = \Big( \frac{X_k}{X_{k'}} \Big)^{1/(k-k')},
\end{equation}
with nonnegative integers, $k \ne k'$, and the moments of the
correlation function
\begin{equation}
 \begin{split}
 X_k &= 
  \frac{1}{k!N_\text{s}} \sum_{i=-\infty}^{\infty} \sum_{\bm{r},\bm{r}'} |i|^k 
  \langle S^z_{0,\bm{r}} S^z_{i,\bm{r}'}
  \rangle \\
  &= \frac{1}{4k!N_\text{s}} 
  \Big\langle \sum_\ell n_\ell^{(0)} m_\ell^k \Big\rangle_\text{MC},
  \end{split}
\end{equation}
where $m_\ell^k \equiv \sum_{i,\bm{r},\alpha} |i|^k
C_\ell(i,\bm{r},\alpha)$.  Note that the moment $X_k$ converges for any
integer $k \ge 0$, since the correlation function decays exponentially
on a finite-width strip even below the critical temperature.  In the
following analysis we use the simplest estimator, $\xi_{1,0} = X_1/X_0$,
since its statistical error is the smallest among all the combinations
$(k,k')$ we examined.

\subsection{Finite-size scaling analysis}

\begin{figure}
 \includefigure{0.48\textwidth}{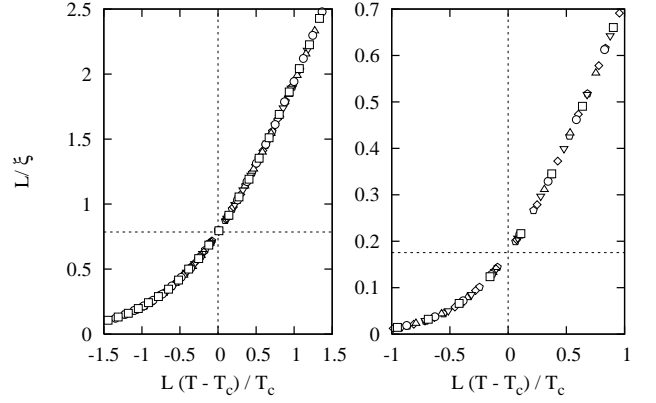}
 \caption{Scaling plots of the correlation length for the $S=1/2$ SQ
 lattices with $J'/J=1$ (left) and $J'/J=0.1$ (right).  System sizes are
 $L=22,24,\ldots,30$ and $L=14,16,\ldots,24$, respectively.  The error
 bar of each point is much smaller than the symbol size.  The critical
 temperature and the critical amplitude are estimated as $T_\text{c} /
 S^2J = 2.2695(3)$ and $L/\xi|_{T=T_\text{c}} = 0.789(3)$ for $J'/J=1$,
 and $T_\text{c} / S^2J = 0.9058(2)$ and $L/\xi|_{T=T_\text{c}} =
 0.176(1)$ for $J'/J=0.1$, respectively }
 \label{fig:plot-2d-xi-scale}
\end{figure}

Near the critical point $T=T_\text{c}$ the correlation length and the
susceptibility of a finite system obey the following finite-size scaling
formulas:
\begin{align}
 L / \xi(L,T) &= \tilde{\xi}(L^{1/\nu} (T-T_\text{c})/T_\text{c}) \ \
 \text{and} \\
 L^{-\gamma/\nu} \chi(L,T) &= \tilde{\chi}(L^{1/\nu}
 (T-T_\text{c})/T_\text{c}),
\end{align}
where $\nu$ and $\gamma$ are the critical exponents of the correlation
length and of the susceptibility respectively, and $\tilde{\xi}(x)$ and
$\tilde{\chi}(x)$ are scaling functions.  In the following analysis we
assume the critical exponents of the Ising universality, i.e., $\nu=1$
and $\gamma/\nu=7/4$ for SQ and SQ+ANNNI, and $\nu=0.6289$ and
$\gamma/\nu=1.9828$ for SC.\cite{FerrenbergL1991}  The critical
temperature is then determined as the point where all the data points
collapse best on a single curve.

As a test of the numerical method explained in this section, we perform
a Monte Carlo simulation and a finite-size scaling analysis for $S=1/2$
SQ lattices with $J'/J = 1$ and 0.1.  In
Fig.~\ref{fig:plot-2d-xi-scale}, we presents scaling plots of the
correlation length.  All the data points collapse excellently on a
single curve in both cases.  The critical temperature is estimated as
$T_\text{c} / S^2J = 2.2695(3)$ and 0.9058(2) for $J'/J = 1$ and 0.1,
respectively, which coincide with the exact values $T_\text{c} / S^2J
\approx 2.26919$ and $0.90588$.

Here we comment that the present lattice geometry $L \times \infty$ is
the same as the one used for the phenomenological renormalization
calculation of the transfer matrix.  As a result, not only the critical
exponents but also the critical amplitude of the correlation length
along the strip is given analytically in the isotropic case ($J'=J$) due
to the conformal invariance of the model at the critical
point.\cite{Cardy1984} We confirm that the critical amplitude
$L/\xi|_{T=T_\text{c}} = 0.789(3)$ for $J'/J=1$
(Fig.~\ref{fig:plot-2d-xi-scale}) agrees with $\pi/4 \approx 0.785$, the
prediction from the conformal field theory.  In anisotropic cases $J'
\ne J$, a nonuniversal prefactor, which represents effective aspect
ratio, is introduced and thus the overall critical amplitude becomes
nonuniversal.  However, this gives us a chance to estimate the
effective aspect ratio at the critical point directly.  For example, we
can estimate the effective aspect ratio immediately as $0.789/0.176 \approx
4.48$ for $J'/J=0.1$.  This method works also for the three-dimensional
case (but does not for the SQ+ANNNI lattice due to the anisotropy
intrinsic to the lattice).  On the other hand, it is practically a hard
task if one works in the standard square or cubic geometry, $L \times
L'$ or $L \times L' \times L''$, for Monte Carlo simulations.  This is
another advantage of the present Monte Carlo method for analyzing
spatial anisotropy precisely, though we will not utilize this quantity
in the remainder of this paper.

\section{Results of Monte Carlo Simulation}
\label{sec:results}

By using the Monte Carlo method and the finite-size-scaling technique
explained in the last section, we estimate the critical temperature of
the SQ, SQ+ANNNI, and SC lattices very carefully for $S=1/2$, 1, and
$3/2$ and $J'/J = 0.001, 0.002, 0.005, \ldots,1$.  The error bar of the
estimates is typically less than 0.01\%.  Such a high accuracy for the
critical temperature is essential for a reliable estimation of the
effective coordination number especially for small $J'/J$, where any
small error in $T_\text{c}$ is amplified vastly due to the exponential
growth of the single-chain susceptibility at low temperatures.  The
results are summarized in Tables~\ref{tab:tc-sq}, \ref{tab:tc-sq-annni},
and \ref{tab:tc-sc}, together with the maximum system size for each
value of $J'/J$.

The present estimates of the critical temperature for the $S=1$ and 3/2
isotropic SQ lattices ($J'/J=1$) can be compared with the previous
high-temperature series expansion study,\cite{ButeraCG2003}
$T_\text{c}/S^2J = 1.69356(1)$ and $1.46144(1)$, respectively.  The
result for the $S=1/2$ isotropic SC lattice also agrees with the
previous Monte Carlo study,\cite{FerrenbergL1991} $T_\text{c}/S^2J =
4.511424(53)$ within the error bar.

\begin{figure}
 \includefigure{0.44\textwidth}{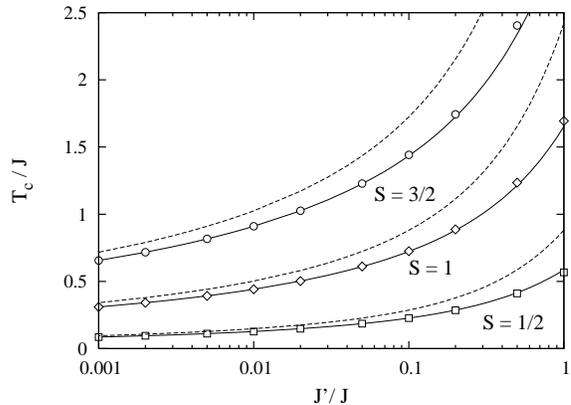}
 \caption{$J'$-dependence of the critical temperature for the SQ lattice
 by the CMF approximation (dashed lines) compared with the exact
 [$S=1/2$ (squares)] and the Monte Carlo [$S=1$ (diamonds) and 3/2
 (circles)] results.  Solid lines denote the CMF results with a
 renormalized coordination number $\zeta = 1$ (see the text for
 details).  The error bar of the data for $S=1$ and 3/2 is much smaller
 than the symbol size.}  \label{fig:tc-log}
\end{figure}

For all the lattices and all the spin sizes we examined, the critical
temperature, normalized by $J$, becomes lower with decreasing $J'/J$, as
expected.  The decrease is found extremely slow.  Even at $J'/J=0.001$
the critical temperature is as high as 10\%--20\% of the isotropic case.
This slow decrease is explained, within the CMF theory, as a reflection
of exponential divergence of the single-chain susceptibility
[Eq.~(\ref{eqn:asym-chi-1d})].

In Fig.~\ref{fig:tc-log}, we plot the critical temperature of the SQ
lattice as a function of $J'/J$ together with the estimates by the CMF
approximation, i.e., the numerical solutions of Eq.~(\ref{eqn:cmf-tc}).
One sees that the CMF theory constantly overestimates the critical
temperature.  This tendency is naturally understood as a systematic
error subsistent in the approximation, introduced by ignoring
fluctuations of order parameter between chains.  Although the deviation
of the CMF results from the correct $T_\text{c}(J')$ becomes smaller
with decreasing $J'/J$, its convergence is logarithmically slow for all
the spin sizes, as demonstrated for $S=1/2$ in Sec.~\ref{sec:intro}.
Similar $J'$-dependence of the critical temperature is observed for the
other lattices.

In Sec.~\ref{sec:intro}, we see that the effective coordination number
for the $S=1/2$ SQ lattice converges to unity instead of 2, which
indicates that the critical temperature in the weak interchain coupling
regime is well-described by a CMF approximation with the renormalized
coordination number, $\zeta=1$.  In Fig.~\ref{fig:tc-log}, we also plot
the results of the CMF approximation with $\zeta=1$ by solid lines, which
amazingly agrees with the true critical temperature in a very wide range
of $J'/J$; even at $J'/J=1$, deviation from the correct value is only
3\%, 2\%, and 4\% for $S=1/2$, 1, and 3/2, respectively, which is much
smaller than those of the conventional CMF theory, 55\%, 43\%, and 39\%.
For $J'/J \lesssim 0.1$, deviation cannot be recognized anymore by the
present scale of the vertical axis in Fig.~\ref{fig:tc-log}.

\begin{table*}
 \caption{Critical temperature and effective coordination number of the
 SQ lattice for $S=1/2$, 1, and 3/2.  The critical temperature for
 $S=1/2$ is obtained from Eq.~(\ref{eqn:onsager}).  The effective
 coordination number for $S=1$ and 3/2 in the $J'/J=0$ limit, presented
 in the bottom row, is estimated by taking an average for
 $J'/J=0.001$, 0.002 and 0.005.}  \label{tab:tc-sq}
 \begin{tabular}{lcllllll}
  \hline
  \hline
  & & \multicolumn{2}{c}{$S=1/2$} & \multicolumn{2}{c}{$S=1$} &
  \multicolumn{2}{c}{$S=3/2$} \\
  \multicolumn{1}{c}{$J'/J$} & \multicolumn{1}{c}{$L_\text{max}$} &
  \multicolumn{1}{c}{$T_\text{c} / S^2J$} & \multicolumn{1}{c}{$\zeta$} &
  \multicolumn{1}{c}{$T_\text{c} / S^2J$} & \multicolumn{1}{c}{$\zeta$} &
  \multicolumn{1}{c}{$T_\text{c} / S^2J$} & \multicolumn{1}{c}{$\zeta$} \\
  \hline
  1     & 72 & 2.26919  & 0.939927 & 1.6934(2)  & 1.0492(2) & 1.4611(1) &
  1.0907(2) \\
  0.5   & 48 & 1.64102  & 0.970162 & 1.2352(1)  & 1.0430(2) & 1.0686(1) &
  1.0731(3) \\
  0.2   & 44 & 1.14156  & 0.989893 & 0.88657(5) & 1.0230(2) & 0.77513(5) &
  1.0393(2) \\
  0.1   & 40 & 0.905883 & 0.995958 & 0.72490(7) & 1.0111(4) & 0.64059(6) &
  1.0191(4) \\
  0.05  & 36 & 0.741313 & 0.998486 & 0.61061(3) & 1.0045(2) & 0.54589(3) &
  1.0082(3) \\
  0.02  & 36 & 0.590681 & 0.999618 & 0.50250(2) & 1.0010(2) & 0.45587(2) &
  1.0021(2) \\
  0.01  & 36 & 0.508926 & 0.999871 & 0.44146(2) & 1.0002(3) & 0.40448(2) &
  1.0004(3) \\
  0.005 & 32 & 0.445462 & 0.999958 & 0.39256(2) & 0.9998(3) & 0.36285(2) &
  0.9999(4) \\
  0.002 & 32 & 0.380981 & 0.999991 & 0.34134(1) & 0.9999(2) & 0.31863(2) &
  0.9997(5) \\
  0.001 & 32 & 0.342657 & 0.999997 & 0.31008(2) & 0.9997(5) & 0.29131(2)
  & 1.0000(6)
  \\
  \hline
  & & & 1 & & 0.99989(6) & & 0.99986(9) \\
  \hline
  \hline
 \end{tabular}

 \caption{Critical temperature and effective coordination number of the
 SQ+ANNNI lattice for $S=1/2$, 1, and 3/2.  The effective coordination
 number in the $J'/J=0$ limit, presented in the bottom row, is
 estimated by taking an average for $J'/J=0.001$, 0.002 and 0.005.}
 \label{tab:tc-sq-annni}
 \begin{tabular}{lcllllll}
  \hline
  \hline
  & & \multicolumn{2}{c}{$S=1/2$} & \multicolumn{2}{c}{$S=1$} &
  \multicolumn{2}{c}{$S=3/2$} \\
  \multicolumn{1}{c}{$J'/J$} & \multicolumn{1}{c}{$L_\text{max}$} &
  \multicolumn{1}{c}{$T_\text{c} / S^2J$} & \multicolumn{1}{c}{$\zeta$} &
  \multicolumn{1}{c}{$T_\text{c} / S^2J$} & \multicolumn{1}{c}{$\zeta$} &
  \multicolumn{1}{c}{$T_\text{c} / S^2J$} & \multicolumn{1}{c}{$\zeta$} \\
  \hline
  1     & 72 & 3.8487(9)  & 2.2889(8) & 2.8064(4)  & 2.5325(6) &
  2.4032(2) & 2.6157(3) \\
  0.5   & 48 & 2.5799(3)  & 2.3766(5) & 1.8790(3)  & 2.5656(8) &
  1.6084(2)  & 2.6223(6) \\
  0.2   & 44 & 1.6451(1)  & 2.4388(3) & 1.2240(1)  & 2.5495(5) &
  1.0548(1)  & 2.5947(6) \\
  0.1   & 40 & 1.2378(1)  & 2.4600(5) & 0.9478(2)  & 2.522(2)  &
  0.8244(1)  & 2.550(1) \\
  0.05  & 36 & 0.97023(6) & 2.4698(5) & 0.76744(5) & 2.4990(6) &
  0.67517(7) & 2.513(1) \\
  0.02  & 36 & 0.73958(5) & 2.4747(6) & 0.60904(4) & 2.4831(7) &
  0.54444(4) & 2.4889(8) \\
  0.01  & 36 & 0.62076(3) & 2.4757(5) & 0.52440(4) & 2.4786(9) &
  0.47414(3) & 2.4810(8) \\
  0.005 & 32 & 0.53183(3) & 2.4751(7) & 0.45878(2) & 2.4768(6) &
  0.41909(2) & 2.4768(7) \\
  0.002 & 32 & 0.44467(2) & 2.4756(6) & 0.39197(1) & 2.4764(4) &
  0.36232(2) & 2.4755(9) \\
  0.001 & 32 & 0.39441(2) & 2.4757(8) & 0.35215(1) & 2.4758(5) &
  0.32803(2) & 2.476(1) \\
  \hline
  & & & 2.4755(2) & & 2.4763(3) & & 2.4763(4) \\
  \hline
  \hline
 \end{tabular}

 \caption{Critical temperature and effective coordination number of the
 SC lattice for $S=1/2$, 1, and 3/2.  The effective coordination number
 in the $J'/J=0$ limit, presented in the bottom row, is estimated
 by taking an average for $J'/J=0.001$, 0.002 and 0.005.}
 \label{tab:tc-sc}
 \begin{tabular}{lcllllll}
  \hline
  \hline
  & & \multicolumn{2}{c}{$S=1/2$} & \multicolumn{2}{c}{$S=1$} &
  \multicolumn{2}{c}{$S=3/2$} \\
  \multicolumn{1}{c}{$J'/J$} & \multicolumn{1}{c}{$L_\text{max}$} &
  \multicolumn{1}{c}{$T_\text{c} / S^2J$} & \multicolumn{1}{c}{$\zeta$} &
  \multicolumn{1}{c}{$T_\text{c} / S^2J$} & \multicolumn{1}{c}{$\zeta$} &
  \multicolumn{1}{c}{$T_\text{c} / S^2J$} & \multicolumn{1}{c}{$\zeta$} \\
  \hline
  1     & 22 & 4.5117(3)  & 2.8962(3) & 3.1966(4)  & 3.0814(6) &
  2.7138(3)  & 3.1435(5) \\
  0.5   & 20 & 2.9297(3)  & 2.9606(5) & 2.0875(1)  & 3.1077(3) &
  1.7746(1)  & 3.1583(3) \\
  0.2   & 20 & 1.8139(1)  & 3.0111(3) & 1.32750(6) & 3.1015(3) &
  1.1378(2)  & 3.137(1)  \\
  0.1   & 20 & 1.34307(7) & 3.0296(4) & 1.01438(5) & 3.0836(4) &
  0.87813(2) & 3.1060(2) \\
  0.05  & 18 & 1.03984(7) & 3.0387(6) & 0.81302(2) & 3.0654(3) &
  0.71228(1) & 3.0787(2) \\
  0.02  & 18 & 0.78297(2) & 3.0434(3) & 0.63900(1) & 3.0514(2) &
  0.56905(2) & 3.0563(5) \\
  0.01  & 18 & 0.65251(1) & 3.0440(2) & 0.54724(1) & 3.0469(3) &
  0.49310(1) & 3.0486(3) \\
  0.005 & 14 & 0.55589(3) & 3.0444(8) & 0.47673(2) & 3.0455(7) &
  0.43420(2) & 3.0459(8) \\
  0.002 & 12 & 0.46200(1) & 3.0447(4) & 0.40544(1) & 3.0445(4) &
  0.37385(2) & 3.044(1)  \\
  0.001 & 10 & 0.40832(2) & 3.0463(8) & 0.36325(3) & 3.046(2) &
  0.33763(3) & 3.045(2)  \\
  \hline
  & & & 3.0449(4) & & 3.0449(3) & & 3.0453(6) \\
  \hline
  \hline
 \end{tabular}
\end{table*}

\begin{figure}
 \includefigure{0.44\textwidth}{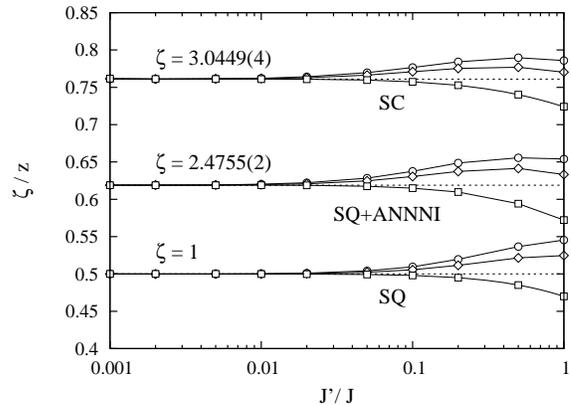}
 \caption{$J'$-dependence
 of the effective coordination number [Eq.~(\ref{eqn:def-zeta})] for the
 SQ, SQ+ANNNI, and SC lattices with $S=1/2$ (squares), 1 (diamonds), and
 3/2 (circles).  The error bar of each point, except for the exact
 results for the $S=1/2$ SQ lattice, is much smaller than the symbol
 size.  Horizontal dashed lines denote the limiting value for each
 lattice evaluated by taking an average of $\zeta(J')$ at $J'=0.001$,
 0.002, and 0.005 for $S=1/2$.}
 \label{fig:zeta}
\end{figure}

To investigate the coordination number reduction in more detail, we next
plot the effective coordination number [Eq.~(\ref{eqn:def-zeta})] as a
function of $J'/J$ in Fig.~\ref{fig:zeta}.  Precise values are also
listed in Tables~\ref{tab:tc-sq}, \ref{tab:tc-sq-annni}, and
\ref{tab:tc-sc}.  One sees in Fig.~\ref{fig:zeta} that the effective
coordination number clearly converges to a finite value very rapidly for
all the lattices and all the spin sizes.  The $S$-independence of the
limiting value is confirmed to an accuracy of 0.01\% (see the bottom row
of Tables~\ref{tab:tc-sq}, \ref{tab:tc-sq-annni}, and \ref{tab:tc-sc}).
We conclude $\zeta$ in the weak interchain coupling limit as 1,
2.4755(2), and 3.0449(4) for the SQ, SQ+ANNNI, and SC lattices,
respectively.  The renormalized coordination number for the SC lattice
is significantly larger than the estimate for the Heisenberg
antiferromagnets,\cite{YasudaTHAKTT2005} $\zeta = 2.78$.  That is, the
renormalized coordination number is different for models with different
spin symmetry.

The rapidness of the convergence of $\zeta(J')$ is quite remarkable.
This makes a sharp contrast to the susceptibility of a single chain.
For example, at $J'/J=0.01$ the critical temperature, normalized by
$S^2J$, for $S=3/2$ ranges between 0.4 and 0.5.  In this temperature
range the single-chain susceptibility deviates from the asymptotic
scaling form by $\approx20\%$ (see the inset of Fig.~\ref{fig:chi-1d}).
On the other hand, the effective coordination number, which is merely
the inverse of the single-chain susceptibility, divided by $J'$, already
converges well to the $J'/J = 0$ limit (Fig.~\ref{fig:zeta}).  This
implies that a cancellation of corrections to the asymptotic scaling
behavior occurs at some deeper level.  In other words, existence of a
more fundamental equation than the modified CMF relation, i.e.,
Eq.~(\ref{eqn:cmf-tc}) with $z$ replaced by $\zeta$, is strongly
suggested.

Although the renormalization of the coordination number is observed
independent of $S$, the same as for the quantum Heisenberg
antiferromagnets reported in Ref.~\onlinecite{YasudaTHAKTT2005}, 
one should however notice that $\zeta$ has different values evidently for the SQ
and SQ+ANNNI lattices, that is, it depends not only on the
dimensionality but also on the real coordination number of the lattice,
$z$.  That means that the ``universality'' we observe in the renormalized
coordination number is somewhat weaker than that for critical exponents,
which depend only on the dimensionality and not on the fine structure of
the lattice, in the usual context of continuous phase transitions.  In
the next section we will unveil the origin of the ``universality'' as well
as the ``nonuniversality'' in the renormalized coordination number for the
present Q1D Ising model.

\section{Mapping to Quantum Ising Model}
\label{sec:mapping}

\subsection{Spin-1/2 case}

In the weak interchain coupling regime, $J'/J \ll 1$, the correlation
length along chains grows much faster than in the perpendicular
directions as the temperature is decreased, and it reaches the
one-dimensional scaling region, described by Eq.~(\ref{eqn:asym-xi-1d}),
well before the correlation in the perpendicular directions starts to
develop over more than a few times of the lattice spacing.  In such a
strongly anisotropic regime, it can be justified to take a continuous
limit only in the chain direction, while keeping the lattice in the
other directions discrete, without introducing any uncontrollable
systematic errors.  In this limit, as is well-known, a $d$-dimensional
classical Ising model is mapped onto a $(d-1)$-dimensional quantum Ising
model in a transverse field and vice versa.\cite{Suzuki1976} In this
section, we first consider the mapping of the $S=1/2$ classical Ising
model.  Generalization for $S \ge 1$ will be separately discussed
shortly.

In order to derive an explicit relation between the parameters of the
classical and the quantum models, we again introduce the transfer matrix
formulation for the $d$-dimensional classical Ising model.  The
partition function of the $S=1/2$ classical Ising model
[Eq.~(\ref{eqn:hamiltonian})] is expressed as
\begin{equation}
 \label{eqn:z-classical}
  Z =\lim_{M\rightarrow\infty} \text{tr} \,
  (\bm{T}\bm{D})^M.
\end{equation}
Here $\bm{T}$ is a $2^{N_\text{s}} \times 2^{N_\text{s}}$ matrix along
chains and is a direct product of local transfer matrices
\begin{equation}
\bm{T} = \bigotimes_\bm{r} e^{J S^{z}_{i,\bm{r}} S^{z}_{i+1,\bm{r}}/T}
 = \bigotimes_\bm{r} \left(
\begin{array}{cc}
e^{J/4T} & e^{-J/4T} \\
e^{-J/4T} & e^{J/4T}
\end{array} \right)
\end{equation}
and $\bm{D}$ is a diagonal matrix representing the Boltzmann weight of
each slice
\begin{equation}
\bm{D} = \exp \Big( \frac{J'}{T} \sum_{\bm{r},\bm{\delta}}
S^{z}_{i,\bm{r}} S^{z}_{i,\bm{r}+\bm{\delta}} \Big).
\end{equation}

On the other hand, the zero-temperature partition function of the
quantum spin-$\frac{1}{2}$ transverse-field Ising model defined on a
$(d-1)$-dimensional lattice is written as
\begin{equation}
 \label{eqn:z-quantum}
 \begin{split}
  Z_\text{q} &= \lim_{T \rightarrow 0} \text{tr} \, e^{-{\cal H}_\text{q}/T} \\
  &= \lim_{\Delta \rightarrow 0} \text{tr} \, e^{-\Delta {\cal
  H}_\text{t}}  e^{-\Delta {\cal H}_\text{d}} e^{-\Delta {\cal
  H}_\text{t}}  e^{-\Delta {\cal H}_\text{d}} \cdots,
 \end{split}
\end{equation}
where the quantum Hamiltonian
\begin{equation}
\label{eqn:tim}
 \begin{split}
 {\cal H}_\text{q} &= {\cal H}_\text{t} + {\cal H}_\text{d} =
  - \Gamma \sum_{\bm{r}}
 \sigma_\bm{r}^x
  - K
 \sum_{\bm{r},\bm{\delta}} \sigma_{\bm{r}}^z
 \sigma_{\bm{r}+\bm{\delta}}^z
 \end{split}
\end{equation}
is defined on a $(d-1)$-dimensional lattice which has the same structure
as a slice of the classical Ising model under consideration.  Here
$\sigma^\alpha_{\bm{r}}$ ($\alpha=x$, $z$) is a Pauli spin operator.

By identifying $\bm{T}$ and $\bm{D}$ in the classical partition
function~(\ref{eqn:z-classical}) with $e^{-\Delta {\cal H}_\text{t}}$
and $e^{-\Delta {\cal H}_\text{d}}$ of the transverse-field Ising
model~(\ref{eqn:z-quantum}) respectively, one can readily obtain the
following relations:
\begin{align}
 \label{eqn:corr-1-1}
 e^{-J/2T} &= \tanh \Delta \Gamma \ \ \text{and} \\
 \label{eqn:corr-1-2}
 \frac{J'}{4T} &= \Delta K
\end{align}
between the parameters of the two models, $(J,J',T)$ and $(K,\Gamma)$, or
\begin{equation}
 e^{-J/2T} = \tanh \frac{J'\Gamma}{4KT} \simeq \frac{J'\Gamma}{4KT}
\end{equation}
by eliminating $\Delta$ in Eqs.~(\ref{eqn:corr-1-1}) and
(\ref{eqn:corr-1-2}).  The last equation is valid in the limit of $T/J$,
$J'/T \rightarrow 0$, and relates the temperature $T$ of the classical
model in this limit to the strength of the transverse field $\Gamma$ of
the quantum model at the ground state.  Especially, the critical points
$T_\text{c}$ of the former at $J' \ll J$
and the quantum critical point
$\Gamma_\text{c}$ of the latter is identified through
\begin{equation}
 \label{eqn:41}
 e^{-J/2T_\text{c}} \simeq \frac{J'\Gamma_\text{c}}{4KT_\text{c}}.
\end{equation}

Now we are ready to calculate the effective coordination
number~(\ref{eqn:def-zeta}) in the weak interchain coupling limit.  From
Eqs.~(\ref{eqn:chi-half}) and (\ref{eqn:41}),
\begin{equation}
 \label{eqn:42}
 \begin{split}
  \zeta(J') &= \frac{1}{J' \chi^\text{1d}(T_\text{c}(J'))}
  = \frac{4T_\text{c}}{J'} e^{-J/2T_\text{c}}
  \\
  & \!\!\!\! \underset{J'\rightarrow0}{\longrightarrow}
  \frac{4T_\text{c}}{J'} \frac{J' \Gamma_\text{c}}{4KT_\text{c}}
  = \frac{\Gamma_\text{c}}{K}.
 \end{split}
\end{equation}
Thus the effective coordination number converges to a finite value in
the $J'=0$ limit and its limiting value is {\em exactly} given by the
quantum critical point, $\Gamma_\text{c}/K$, of the corresponding
$(d-1)$-dimensional quantum Ising model.

One can indeed confirm that the analytic and numerical results for the
renormalized coordination number, $\zeta = 1$ and 3.0449(4) for the SQ
and SC lattices, in the last section agrees completely with the quantum
critical point of the transverse-field Ising model, $\Gamma_\text{c}/K =
1$ (Ref.~\onlinecite{Pfeuty1970}) and 3.0440(7) (Ref.~\onlinecite{HeHO1990}) or
3.0450(2) (Ref.~\onlinecite{Hamer2000}) on the single chain and the isotropic square
lattice, respectively.  Note that in the argument presented above, we do
not assume any exact solutions except for the susceptibility of the
genuinely one-dimensional model.  Therefore Eq.~(\ref{eqn:42}) applies
to arbitrary Q1D Ising models in any dimensions.  (For example, it
applies even to the models with frustration in the transverse
directions.)

\subsection{General spin cases}

One may expect that a spin-$S$ classical Ising model could be similarly
mapped onto a quantum Ising model {\em with the identical spin size}.
However, this is not the case.  Such a naive mapping, albeit formally
possible, gives no physically meaningful weak interchain coupling limit.

To see the difficulty for $S\ge1$ more precisely, we consider the $S=1$
case as an example.  The $S=1$ local transfer matrix along chains is
given by
\begin{equation}
\bm{T}_\bm{r} = e^{JS_{i,\bm{r}}^zS_{i+1,\bm{r}}^z/T} = \left(
\begin{array}{ccc}
e^{J/T} & 1 & e^{-J/T} \\
1 & 1 & 1 \\
e^{-J/T} & 1 & e^{J/T}
\end{array} \right).
\end{equation}
If this local transfer matrix is fit by an imaginary-time propagator
$\exp(-\Delta {\cal H}_\text{t,\bm{r}})$ by the following $S=1$ quantum
Hamiltonian:
\begin{equation}
{\cal H}_\text{t,\bm{r}} = -\Gamma \tau^x_{\bm{r}} - D (\tau^z_{\bm {r}})^2 
 -E (\tau^x_{\bm {r}})^2,
\end{equation}
where the $\tau^\alpha_{\bm{r}}$ ($\alpha=x$, $z$) is a spin-1 operator,
$\Gamma$ the transverse field, and $D$ and $E$ are the longitudinal and
transverse crystal fields, respectively, we obtain
\begin{align}
 \Delta \Gamma &\simeq \frac{J}{\sqrt{2} T} e^{-J/T}, \\
 \Delta D &\simeq J/T, \ \ \text{and} \\
 \Delta E &\simeq - 4J/T
\end{align}
up to the leading order in $J/T$.  In the last equations, however, one
should notice that $\Delta \Gamma$ vanishes for $T \ll J$ (so that
$\Gamma$ itself can remain finite), whereas both $D$ and $|E|$
diverge even after being multiplied by $\Delta$.  This desperate asymptotic
behavior of the parameters of the mapped quantum system is due to the
vanishing diagonal weight for the $S^z=0$ state in the transfer matrix
$\bm{T}_\bm{r}$.

Although for $S \ge 1$ the weight of the states with $|S^z| < S$ becomes
exponentially small at low temperatures, disregarding such matrix
elements in the transfer matrix will mislead us into an incorrect
quantum spin Hamiltonian.  In order to handle the anisotropic limit of
the high-spin models with care, we introduce a unitary transformation
defined below.

As mentioned in Sec.~\ref{sec:model}, only the two largest eigenvalues
$\lambda_1$ and $\lambda_2$ of the local transfer matrix $\bm{T}_\bm{r}$
are dominant at low temperatures, and the other $2S-1$ eigenvalues are
exponentially small in comparison with $\lambda_1$ and $\lambda_2$.  Now
we introduce two column vectors $v_1$ and $v_2$, which are linear
combinations of the eigenvectors $u_1$ and $u_2$ associated with
$\lambda_1$ and $\lambda_2$, respectively,
\begin{align}
 v_1 &= (u_1 + u_2) / \sqrt{2} \ \ \text{and} \\
 v_2 &= (u_1 - u_2) / \sqrt{2}.
\end{align}
Clearly $v_1$ and $v_2$ are orthogonal with each other by definition and
furthermore they are interchanged by the spin inversion transformation,
since $u_1$ ($u_2$) is symmetric (antisymmetric) under the
transformation.  This implies that one can interpret $v_1$ and $v_2$ as
``wave functions'' of a pseudo $S=1/2$ quantum spin.  Indeed, by a local
unitary transformation by a matrix
$\bm{V}_\bm{r}=(v_1,v_2,v_3,\ldots,v_{2S+1})$ with $v_1$ and $v_2$
introduced above and $v_i=u_i$ for $i \ge 3$, the local transfer matrix
at low temperatures $T \ll J$ is reduced to a $2 \times 2$ matrix
\begin{equation}
\begin{split}
 \bm{V}_\bm{r}^{-1} {\bm T}_{\bm r} \bm{V}_\bm{r} &= 
 \begin{pmatrix}
  \lambda_+ & \lambda_- & & & 0 \\
  \lambda_- & \lambda_+ & & & \\
  & & \lambda_3 & & \\
  & & & \ddots & \\
  0 & & & & \lambda_{2S+1}
 \end{pmatrix} \\
 & \simeq
 \begin{pmatrix}
  \lambda_+ & \lambda_- & & & 0 \\
  \lambda_- & \lambda_+ & & & \\
  & & 0 & & \\
  & & & \ddots & \\
  0 & & & & 0
 \end{pmatrix},
 \end{split}
\end{equation}
where $\lambda_\pm = (\lambda_1\pm\lambda_2)/2$.  The whole transfer
matrix $\bm{T} = \otimes_\bm{r} \bm{T}_\bm{r}$ of $(2S+1)^{N_\text{s}}
\times (2S+1)^{N_\text{s}}$ is also reduced to a $2^{N_\text{s}} \times
2^{N_\text{s}}$ matrix by the global unitary transformation using
$\bm{V} = \otimes_\bm{r} \bm{V}_\bm{r}$.  We transform the diagonal
matrix $\bm{D}$ at the same time to keep the partition
function~(\ref{eqn:z-classical}) invariant.  Using
Eqs.~(\ref{eqn:14})--(\ref{eqn:17}), relevant matrix elements of
$\bm{D}$ in this new basis are calculated as
\begin{equation}
\begin{split}
 & (\bm{V}_\bm{r} \otimes \bm{V}_{\bm{r}+\bm{\delta}})^{-1}
 \bm{D}_{\bm{r},\bm{r}+\bm{\delta}} (\bm{V}_\bm{r} \otimes
 \bm{V}_{\bm{r}+\bm{\delta}}) \\
 & \qquad \simeq
 \begin{pmatrix}
  e^{S^2J'/T} & & & 0 & \\
  & \!\!\!\! e^{-S^2J'/T} & & & \\
  & & \!\!\!\! e^{-S^2J'/T} & & \\
  0 & & & \!\!\!\! e^{S^2J'/T} & \\
  & & & & \!\!\!\! \ddots
 \end{pmatrix}.
\end{split}
\end{equation}

\begin{figure}
 \includefigure{0.44\textwidth}{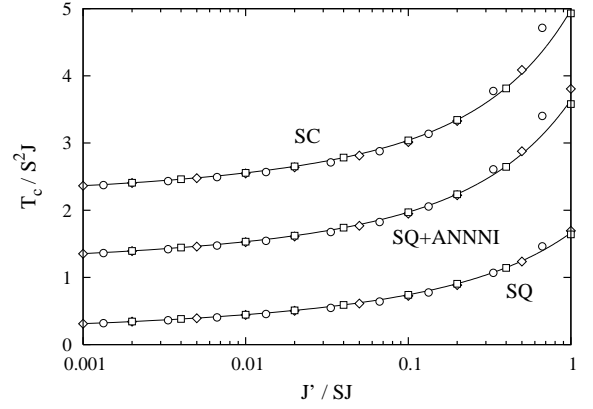}
 \caption{Scaling plot
 of the critical temperature for the SQ, SQ+ANNNI, and SC lattices with
 $S=1/2$ (squares), 1 (diamonds), and 3/2 (circles).  Solid lines denote
 scaling function, $2 y e^{-2/y} = C x$, with $x = J'/SJ$, $y =
 T_\text{c}/S^2J$, and $C = \lim_{J'\rightarrow0} \zeta(J')$ [$\approx1$,
 2.4755, and 3.0449 for SQ, SQ+ANNNI, and SC, respectively].  The error
 bar of each point is much smaller than the symbol size.  The data for
 the SQ+ANNNI and SC lattices are shifted upwards by 1 and 2,
 respectively, for visibility.}  \label{fig:tc-scale}
\end{figure}

The effective partition function of the spin-$S$ classical Ising model
in the weak interchain coupling limit thus becomes equivalent to a
spin-1/2 partition function, and then is mapped onto a spin-1/2 quantum
Ising model, instead of spin-$S$, as the same as the $S=1/2$ classical
Ising model.  Noticing that
\begin{align}
 \begin{pmatrix}
  \lambda_+ & \lambda_- \\
  \lambda_- & \lambda_+
 \end{pmatrix}
 &= \exp 
 \begin{pmatrix}
  \log (\lambda_+ - \lambda_-) & \frac{1}{2} \log \frac{\lambda_+ +
  \lambda_-}{\lambda_+ - \lambda_-} \\
  \frac{1}{2} \log \frac{\lambda_+ + \lambda_-}{\lambda_+ - \lambda_-} &
  \log (\lambda_+ - \lambda_-)
 \end{pmatrix} \\
 &= \exp 
 \begin{pmatrix}
  \log \lambda_1 & \frac{1}{2} \log \lambda_1/\lambda_2 \\
  \frac{1}{2} \log \lambda_1/\lambda_2 & \log \lambda_1
 \end{pmatrix},
\end{align}
one can explicitly write down the correspondence between the spin-$S$
classical Ising model and the spin-1/2 quantum transverse-field Ising
model as
\begin{align}
 \log \frac{\lambda_1(T)}{\lambda_2(T)} &\equiv 1 / \xi_\text{1d}(T) = 2
 \Delta \Gamma \ \ \text{and} \\ 
 \frac{S^2 J'}{T} & = \Delta K
\end{align}
or
\begin{align}
 \label{eqn:57}
 1/\xi_\text{1d}(T) &= \frac{2 S^2 J' \Gamma}{KT}
\end{align}
for $J' \ll T \ll J$.  Furthermore, using the low-temperature
asymptotic form for the correlation length of a single chain
[Eq.~(\ref{eqn:asym-xi-1d})], we obtain a relation between the classical
and the quantum critical point parameters as
\begin{equation}
 \label{eqn:58}
 e^{-2S^2J/T_\text{c}} \simeq \frac{SJ'\Gamma_\text{c}}{2T_\text{c}K},
\end{equation}
which reduces to Eq.~(\ref{eqn:41}) for $S=1/2$.
The last equation suggests that two dimensionless variables, $x \equiv J'/SJ$
and $y \equiv T_\text{c}/S^2J$, satisfy a scaling relation,
$2ye^{-2/y} = C x$ with an $S$-independent constant $C=
\Gamma_\text{c}/K$.  In Fig.~\ref{fig:tc-scale}, this scaling relation
is tested explicitly by using the Monte Carlo results in the last
section.  For each lattice the data in the weak interchain coupling
regime $J'/SJ \lesssim 0.1$ collapse on a single curve as expected from
the scaling form~(\ref{eqn:58}).

Finally, using Eq.~(\ref{eqn:asym-chi-1d}), the effective coordination
number in the $J'/J=0$ limit is obtained as $C = \Gamma_\text{c}/K$,
irrespective of the spin size, which is a proof of the $S$-independent
renormalization of the coordination number suggested by the Monte Carlo
simulation in the last section.  One should note the nontrivial factor
$1/S$ in the scaling variable $x = J'/SJ$, which cancels with the same
factor in the scaling form for the single-chain susceptibility
[Eq.~(\ref{eqn:asym-chi-1d})] to make the effective coordination number
independent of $S$.  We also emphasize that our final result,
$\zeta(J')\rightarrow \Gamma_\text{c}/K$, does not depend on the scaling
conjecture for the susceptibility and correlation length of a single
chain; it can be deduced more directly from Eqs.~(\ref{eqn:20}) and
(\ref{eqn:57}) instead.

\section{Summary and Discussion}
\label{sec:summary}

In this paper we presented our detailed analysis on the critical
temperature and the effective coordination number of the Q1D
general-spin Ising ferromagnets.  The critical temperature is estimated
down to $J'/J = 0.001$ with high accuracy by the cluster Monte Carlo
method performed on infinite-length strips.  It is demonstrated that as
the interchain coupling is decreased the effective coordination number
converges to a finite value $\zeta$, which is definitely less than the
coordination number $z$, quite rapidly.  Furthermore, we show
unambiguously that the renormalized coordination number is independent
of the spin size $S$.  This is the same as the observation for the Q1D
and Q2D quantum Heisenberg antiferromagnets, reported in
Ref.~\onlinecite{YasudaTHAKTT2005}, and thus is expected to be observed
widely in other classes of quasi-low-dimensional magnets.

We explain the origin of the renormalization of the effective
coordination number by considering a mapping to quantum Ising models,
where the spin-$S$ classical Ising model is rigorously shown to be
mapped onto a spin-$1/2$ quantum transverse-field Ising model,
irrespective of the spin size $S$, in the weak interchain coupling
limit, and the effective coordination number is given by the quantum
critical point $\Gamma_\text{c}/K$ of the mapped quantum Ising model.
It should be noticed that in the scaling relation of the
critical temperature [Eq.~(\ref{eqn:58})], a nontrivial factor $1/S$
appears, which cannot be deduced by a naive dimensional analysis.
This factor cancels out with another nontrivial factor in the
single-chain susceptibility [Eq.~(\ref{eqn:asym-chi-1d})], and then
results in an $S$-independent scaling relation.  Another
remarkable fact to be noticed is that corrections to the asymptotic
value in the effective coordination number are much smaller than those in
the single-chain susceptibility itself, which suggests existence of a
more fundamental theory, describing the critical temperature of strongly
anisotropic magnets, than the modified CMF relation,
Eq.~(\ref{eqn:cmf-tc}) combined with $z=\zeta$.

The renormalized coordination number is, however, found to be
nonuniversal with respect to the lattice structure.  This is natural
from the viewpoint of the mapped quantum Ising model, where the critical
transverse field of a quantum critical point cannot be a universal
quantity; it depends on the fine lattice structure as well as on the
details of interaction.  Thus an introduction of a weak further neighbor
interaction or an anisotropy in slices, for example, will modify the
renormalized coordination number.  In this sense, the ``universality,''
discussed in the present paper, is different from that asserted for
critical exponents in usual continuous phase transitions and thus
inexact terminologically.

Finally, we emphasize again that the weak interchain coupling limit of
the Q1D classical Ising model is {\em not the weak coupling limit\,}; it
is still in the strongly correlated regime, i.e., in the vicinity of a
quantum critical point, of the mapped quantum Ising model.  This
explains why the CMF approximation remains inaccurate even in this
limit.  We expect a similar argument will apply to the Q1D and Q2D
quantum Heisenberg models.  It is unclear and remains as an open
problem, however, whether it is possible to generalize the present
mapping approach to the quantum Heisenberg models, where a continuous
axis, the imaginary-time axis, exists intrinsically, and therefore
extra care should be necessary to introduce another continuous axis
representing a strong spatial anisotropy.

\section*{Acknowledgments}

The present author thanks C. Yasuda for stimulating discussions and comments.
The numerical simulations presented in this paper have been done by
using the facility of the Supercomputer Center, Institute for Solid
State Physics, University of Tokyo.  The program used in the present
simulations is based on the ALPS libraries.\cite{ALPSweb, ALPS2005}
Support by Grant-in-Aid for Scientific Research Program (No.~15740232
and No.~18540369) from the Japan Society for the Promotion of Science,
and also by NAREGI Nanoscience Project, Ministry of Education Culture,
Sports, Science, and Technology, Japan is gratefully acknowledged.


\end{document}